\documentclass[preprint,showpacs,preprintnumbers,amsmath,amssymb,superscriptaddress]{revtex4}

\usepackage{chemformula} 
\usepackage[T1]{fontenc} 
\usepackage{amssymb}
\usepackage{amsmath}

\usepackage{hyperref}

\newcommand*\VEC[1]{\boldsymbol{#1}}

\newcommand{\op}[1]{\hat{#1}}

\begin{document}

\author{Ping-Yuan Lo}
\affiliation{Department of Electrophysics, National Chiao Tung University, Hsinchu 300, Taiwan}

\author{Guan-Hao Peng}
\affiliation{Department of Electrophysics, National Chiao Tung University, Hsinchu 300, Taiwan}

\author{Wei-Hua Li}
\affiliation{Department of Electrophysics, National Chiao Tung University, Hsinchu 300, Taiwan}

\author{Yi Yang}
\affiliation{Department of Electrophysics, National Chiao Tung University, Hsinchu 300, Taiwan}

\author{Shun-Jen Cheng}
\affiliation{Department of Electrophysics, National Chiao Tung University, Hsinchu 300, Taiwan}
\email{sjcheng@mail.nctu.edu.tw}

\title{Symmetry governed valley-pseudospin textures of the full-zone excitonic bands of transition-metal dichalcogenide monolayers}







\begin{abstract}
Preserving a high degree of valley polarization of excitons in photo-excited transition-metal dichalcogenide monolayers (TMD-MLs) is desirable for the valley-based photonic applications, but widely recognized as a hard task hindered by the intrinsic electron-hole exchange interaction.
In this study, we present a comprehensive investigation of valley-polarized finite-momentum excitons in WSe$_2$-MLs over the entire Brillouin zone by solving the density-functional-theory(DFT)-based Bethe-Salpeter equation (BSE) under the guidance of symmetry analysis.
We reveal that finite-momentum excitons are actually in general well immune from the exchange-induced valley depolarization, except for those with specific exciton momenta directionally coincide with the axes associated with the $3\sigma_v$ and $3C_2'$ symmetries in TMD-MLs.
Governed by the symmetries, the valley pseudo-spin texture of the full-zone exciton band in the momentum space is locally featured by individual skyrmion-like structures where highly valley-polarized finite-momentum exciton states are centred. Remarkably, we show that the high degrees of valley polarizations of the finite-momentum exciton states are excellently well transferable to the optical polarizations in the resulting phonon-assisted photo-luminescences, suggesting the prospective usefulness of those inter-valley excitons in valley-based photonics.
\end{abstract}

\maketitle


\noindent
\emph{Introduction---} Transition-metal dichalcogenide monolayers (TMD-MLs) have drawn a broad interest in recent years because of the intriguing spin-valley-coupled characteristics in the electronic and excitonic structures. \cite{TFHeinz2010, WYao2012a, WYao2014a}
As a massive Dirac material, the band structure of a TMD-ML is characterized by two distinctive gapped valleys locating at the $K$ and $K'$ corners of the first Brillouin zone (BZ) in the momentum space that follow the distinct optical selection rules and allow for the valley-selective excitation and coherent manipulation.\cite{WYao2014b, WYao2012b, JShan2018a}
Those spin- and valley-resolvable characteristics in the excitonic structure of TMD-MLs serve as a prospective base of spin- and valley-based photonics, as long as a high degree of the excitonic valley polarization can be well maintained in the materials.\cite{GWang2015,Smolenski2016}

However, in reality, the valley-polarization of a bright exciton in a TMD-ML is very likely to be depolarized by the electron-hole (\emph{e-h}) exchange interaction which intrinsically couples two interband excitations in distinct valleys. \cite{WYao2014b, MWWu2014, AHMacDonald2016, GWang2014, Glazov2015, selig2020suppression}
Despite the weak coupling strength of meV scale, the {\it e-h} exchange interaction can intermix the distinct spin-like exchange-free exciton states {\it completely} whenever the exchange-free states possess the same momentum and same energy, which together facilitate the resonant coupling driven by the momentum-conserving exchange interaction.\cite{WYao2014b, SGLouie2015a}
Such a resonant inter-valley coupling exists in the bright exciton with nearly vanishing momentum \cite{WYao2014b} but, without the sustain from certain symmetries, is not necessary to be held by a generic exciton with finite exciton wave vector, $\VEC{k}_{ex}\neq 0$.

In spite of violating the optical selection rules, the finite-momentum exciton states of TMD-MLs have been observabed in advanced optical spectroscopies and shown essentially involved in various optical phenomena including the phonon- or defect-assisted luminescences, \cite{SFShi2019, CHLui2020, WYao2020a, EMalic2020a, CHLui2019a} photoluminescence excitations, \cite{bao2020probing} formation of multi-exciton complexes, \cite{chen2018coulomb} the boost of near-field energy transfers, \cite{HZhu2019a} 
and the optical responses in electron energy loss spectroscopy. \cite{Koitzsch2019, KSuenaga2020}
 To understand the exciton physics in TMD-MLs comprehensively, it is demanded to acquire the sufficient information of the complete exciton band structures over the entire BZ, which however has not been fully revealed so far. \cite{AHMacDonald2015a,Deilmann2019,PHawrylak2020}

In this Letter, we present theoretical and computational studies of the full-zone exciton bands of WSe$_2$-MLs by numerically solving the DFT-based BSE for neutral exciton under the guidance of symmetry analysis.\cite{SJCheng2019}
The quantum nature of the valley-polarized exciton states evolved with varying the exciton momentum over the BZ is visualized by the valley pseudo-spin texture, where the skyrmion-like structures surrounding highly valley-polarized exciton states are revealed. Interestingly, we find that the near-unity valley-polarizations of those inter-valley finite momentum exciton states are excellently well transferrable to the optical polarization through the indirect photo-luminescences (PLs) assited by phonons. \cite{SFShi2019, CHLui2020, WYao2020a}

\noindent
\emph{Theoretical analysis and numerical methodology---} We begin with the exciton states with the well-defined center-of-mass wave vector $\VEC{k} _{ex}$,
$\left| S , \VEC{k} _{ex} \right\rangle = \frac{1}{\sqrt{\mathcal{A}}} \sum _{v c \VEC{k}} A _{S ,\VEC{k} _{ex}} \!\! \left( v c \VEC{k} \right) \hat{c} _{c, \VEC{k} + \VEC{k} _{ex}} ^{\dagger} \hat{h} _{v, -\VEC{k}} ^{\dagger} | GS \rangle $
, written as a linear combination of the configurations of the electron-hole ({\it e-h} ) pairs, $\hat{c} _{c, \VEC{k} + \VEC{k} _{ex}} ^{\dagger} \hat{h} _{v, -\VEC{k}} ^{\dagger} | GS \rangle $,
where the particle operator $\hat{c}_{c,\VEC{k}}^{\dagger}$ ($\hat{h}_{v,-\VEC{k}}^{\dagger}$) is defined to create the electron (hole) of the wave vector $\VEC{k}$ ($-\VEC{k}$) in the conduction band $c$ (corresponding to the missing state at $\VEC{k}$ in the filled valence band $v$) from the ground state of the system with the fully filled valence bands $| GS \rangle$,
$S$ is the index of exciton band, $A_{S, \VEC{k} _{ex}} \!\! \left( v c \VEC{k} \right)$ is the amplitude of the {\it e-h} configuration $\hat{c} _{c, \VEC{k} + \VEC{k} _{ex}} ^{\dagger} \hat{h} _{v, -\VEC{k}} ^{\dagger} | GS \rangle$ in the exciton state, and $\mathcal{A}$ is the area of the two-dimensional (2D) material.
Throughout this work, our study is focussed on the spin-like exciton states with the same particle spin in the $c$- and $v$-bands that can be luminescent under proper assistances of phonon- or defect-scatterings.
The exciton wave function in the reciprocal $\VEC{k}$-space, $A _{S, \VEC{k} _{ex}} \!\! \left( v c \VEC{k} \right)$, follows the Bethe-Salpeter equation that reads \cite{LJSham1966, LJSham1980, SGLouie1998, AHMacDonald2015a, Deilmann2019, SJCheng2019, PHawrylak2020, vasconcelos2018dark}
\begin{align}\label{eqn:BSE}
  &\left[ \epsilon _{c, \VEC{k} + \VEC{k} _{ex}} - \epsilon _{v, \VEC{k}} - E _{S, \VEC{k} _{ex}} ^{X} \right] A _{S, \VEC{k} _{ex}} \!\! \left( v c \VEC{k} \right) \notag \\
  &+ \sum _{v ^{\prime} c ^{\prime} \VEC{k} ^{\prime}} U _{\VEC{k} _{ex}} \!\! \left( v c \VEC{k} , v ^{\prime} c ^{\prime} \VEC{k} ^{\prime} \right) A _{S, \VEC{k} _{ex}} \!\! \left( v ^{\prime} c ^{\prime} \VEC{k} ^{\prime} \right) = 0 ,
\end{align}
where $E _{S, \VEC{k}_{ex}} ^{X}$ is the eigen energy of the exciton state, the first two terms on the left hand side contain the kinetic energies of the electron and hole in a free {\it e-h} pair, $\epsilon_{c,\VEC{k}+\VEC{k}_{ex}}$ and $(-\epsilon_{v,\VEC{k}})$, and the last term is associated with the kernel of {\it e-h} Coulomb interaction that consists of the screened {\it e-h} direct interaction and the {\it e-h} exchange one, $U_{\VEC{k} _{ex}}=- V_{\VEC{k} _{ex}} ^{d} + V_{\VEC{k} _{ex}} ^{x}$. The explicit definitions of $V_{\VEC{k} _{ex}} ^{d}$ and $V_{\VEC{k} _{ex}} ^{x}$ in terms of the Bloch wave functions are given in Supplementary Materials. \cite{Supplement} The screening in the {\it e-h} Coulomb interaction for an exciton in the 2D material \cite{Supplement} is considered on the base of Keldysh formalism. \cite{LVKeldysh1979, PCudazzo2011, AHMacDonald2015a, TCBerkelbach2013, AVStier2018, ERidolfi2018, MLTrolle2017} 

\begin{figure}[t]
\includegraphics[width=0.9\columnwidth]{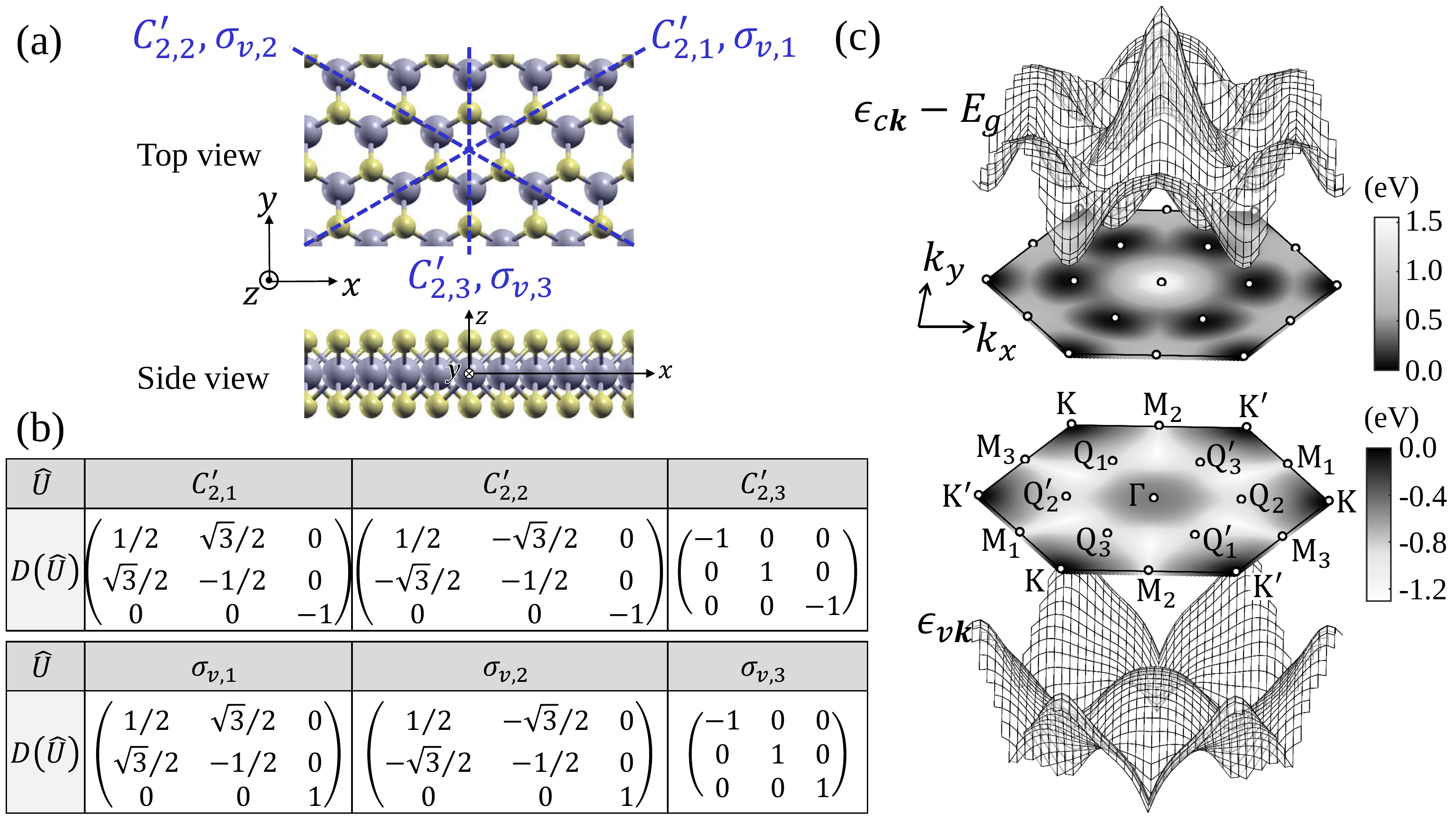}
\caption{(a) Top and side views of the crystalline lattice structure of a TMD-ML with $D_{3h}$ symmetry, where the axes of the $C_{2,i}^{'}$ rotational symmetry and $\sigma_{v,i}$ mirror symmetry relevant to the symmetry analysis are indicated by the dashed lines. (b) List of the matrices of the $C_{2,i}^{'}$ and $\sigma_{v,i}$ symmetry operators in the Cartesian coordinate. (c) Contour plots of the DFT-calculated lowest conduction (top) and highest valence bands (bottom) of a WSe$_2$-ML over the first BZ with the indication of the high symmetry points, where $E_{g}$ is used to indicate the energy gap of a WSe$_2$-ML.} 
\label{Fig1}
\end{figure}

In this work, we follow the developed methodology in Ref.~\cite{SJCheng2019} to set up the BSE theory for the exciton studies on the first principles base. In the approach, the BSE is formulated in the Wannier tight binding scheme \cite{kosmider2013large,scharf2016excitonic,lado2016landau} established by means of wannierization of the DFT-calculated Bloch wave functions.\cite{Wannier90a, Wannier90b, Supplement}
 Figure~\ref{Fig1}c presents the DFT-calculated lowest conduction and topmost valence band of a WSe$_2$-ML over the first BZ by using the first principles VASP package \cite{GKresse1996} with the use of the Perdew–Burke-Ernzerhof (PBE) exchange-correlation functional. \cite{PBE1996} To solve the exciton band structures, the DFT-based BSE is discretized with the $\VEC{k}$ mesh grids compatible to the symmetries of $D_{3h}$ TMD-MLs \cite{PHawrylak2020} and solved numerically by means of direct diagonalization.

It is well established that the direct Coulomb interaction makes the predominant contribution to the large binding energy of exciton at the scale of hundreds of meV in a TMD-ML, but has no direct effect on the coupling between the distinct spin-like exciton states belonging to the opposite valleys. \cite{SGLouie2013a, TFHeinz2014a}
By contrast, the momentum-conserving {\it e-h} exchange interaction at the scale of merely meV could couple the distinct valley-excitons which possess the same exciton momentum $\VEC{k}_{ex}$ and leads to the unwanted valley-depolarization of exciton in the valley-based applications. \cite{MWWu2014, AHMacDonald2016}
In fact, the valley-intermixing of excitons is significant only when the exchange-free valley-polarized states are degenerate or nearly degenerate with the splitting much smaller than the meV-scaled \emph{e-h} exchange interaction.

Thus, we proceed with the symmetry analysis for the free {\it e-h}-pair excitations over the BZ to predict the valley-degenerate exchange-free exciton states of $D_{3h}$ TMD-MLs, where $D_{3h}=\{E,C_{3},C_{3}^{-1},\sigma_{h},S_{3},S_{3}^{-1},C_{2,1}^{\prime},C_{2,2}^{\prime},C_{2,3}^{\prime},\sigma_{v,1},\sigma_{v,2},\sigma_{v,3}\}$.\cite{CRobert2017}
Consider two distinct {\it e-h} pair states with the same $\VEC{k} _{ex}$ excited from the different valence states at $\VEC{k}$ and $\VEC{k} ^{\prime}$, respectively, the degeneracy is formed if
$\epsilon_{c, {\VEC{k} + \VEC{k} _{ex}}} - \epsilon _{v, \VEC{k}} = \epsilon _{c, \VEC{k} ^{\prime} + \VEC{k} _{ex}} - \epsilon _{v, \VEC{k} ^{\prime}} $, which can generally hold only if
$\epsilon_{v, \VEC{k}} = \epsilon _{v, \VEC{k} ^{\prime}}$ and $\epsilon _{c, \VEC{k} + \VEC{k} _{ex}} = \epsilon _{c, \VEC{k} ^{\prime} + \VEC{k} _{ex}}$.
From the theory of group representations, the above two equations hold when the space group symmetry of the TMD-ML satisfies the both equations, $\VEC{k} ^{\prime} = \hat{U} \VEC{k}$ and $\VEC{k} ^{\prime} + \VEC{k} _{ex} = \hat{U} \left( \VEC{k} + \VEC{k} _{ex} \right)$, for any symmetry operator $\hat{U} \in D _{3h}$.
Accordingly, we find the criterion for the formation of valley-degeneracy of two distinct {\it e-h} pairs carrying the same $\VEC{k}_{ex}$, i.e.
\begin{equation}
\VEC{k}_{ex} = \hat{U} \VEC{k}_{ex} \, .
\label{exc_deg}
\end{equation}
The matrix representations of $\hat{U} \in D _{3h}$ in the Cartesian coordinate are denoted by $D ( \hat{U} )$ and explicitly given in Supplementary Materials. \cite{Supplement} Figure~\ref{Fig1}b lists $D ( \hat{U} )$ for $\hat{U} \in C_{2,i}^{'}$ and $\sigma_{v,i}$.

Applying all the symmetry operators $\hat{U}$ onto Eq.~(\ref{exc_deg}) for all $\VEC{k} _{ex} \in $ BZ, one can show that distinct \emph{e-h} pair states with the common $\VEC{k} _{ex}$ could be valley-degenerate only if $\VEC{k} _{ex}$ lies along the lines connecting the $\Gamma _{ex}$ and $M _{ex,i}$ points, i.e. the axes associated with the $3\sigma_v$ and $3C_2'$ symmetries as shown in Fig.~\ref{Fig1}a.
This predicts the impactive valley depolarization only happening in the exciton states with the specific $\VEC{k} _{ex}$ in coincidence with the $\overline{\Gamma_{ex}M_{ex,i}}$ paths, including the commonly known bright exciton around the the $\Gamma_{ex}$ point. \cite{MWWu2014, AHMacDonald2016}



\begin{figure}[t]
\includegraphics[width=0.9\columnwidth]{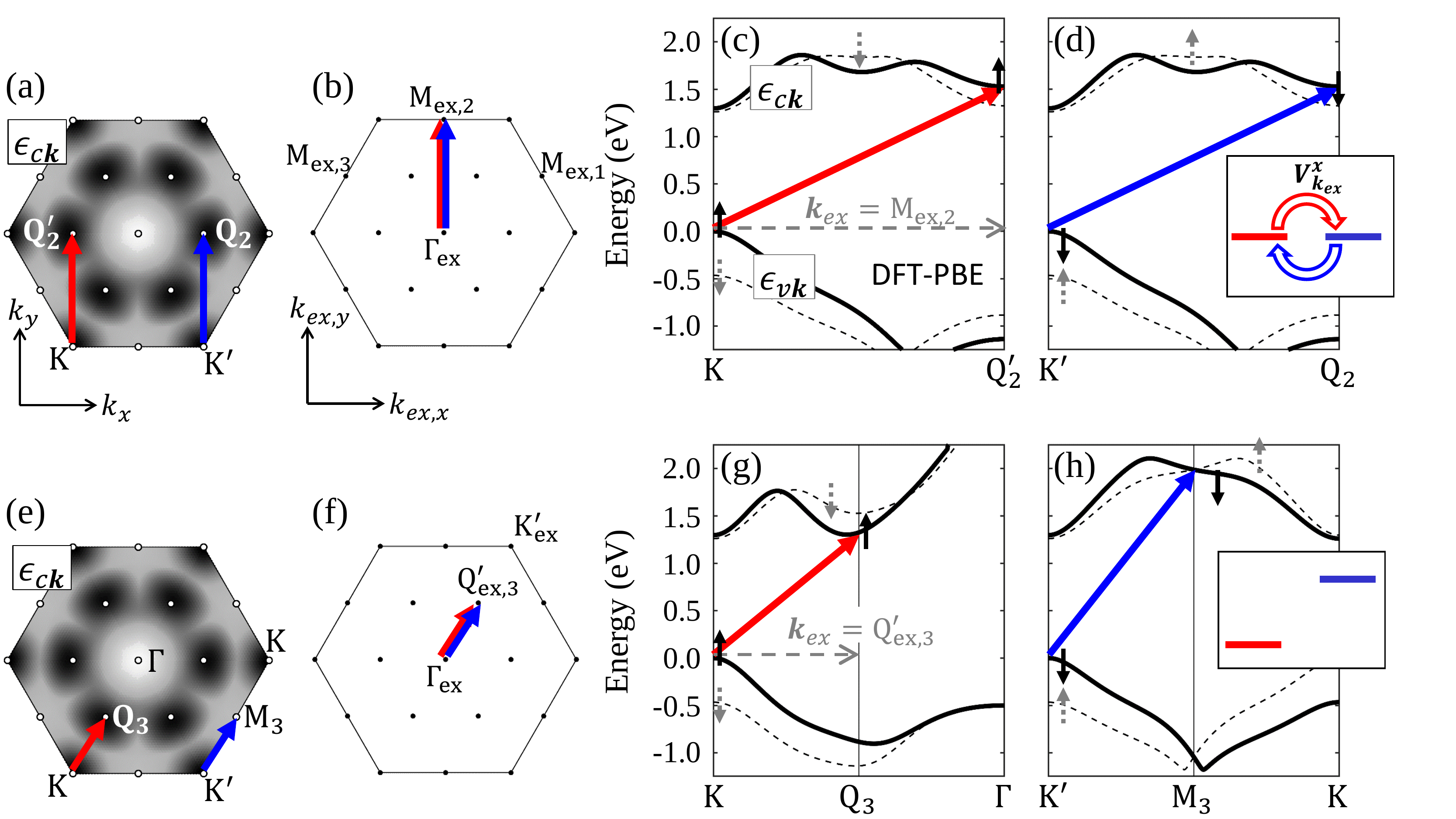}
\caption{ (a) Two distinct electron-hole pair excitations with the same $\VEC{k}_{ex}$ (directed along $\overline{\Gamma_{ex} M_{ex,2}}$), excited from the valence $K$ to the conduction $Q_2^{'}$ valleys (red arrow line) and from the valence $K^{'}$ to the $Q_2$ conduction valleys (blue arrow line), respectively. (b) The BZ in the exciton-momentum space with the indication of the two excitations. (c) [(d)] The DFT-calculated quasi-particle band dispersion along the momentum direction of the first [second] excitation. Arrow lines indicates the spin-like excitations. (e)-(h): same as the graphs of (a)-(d) but for another set of distinct excitations with the common $\VEC{k}_{ex}\parallel \overline{\Gamma_{ex} Q_{ex,3}^{'}}$. The inset of (d) [(h)] illustrates the existence [absence] of the resonant inter-valley coupling between the two degenerate [non-degenerate] valley-excitations in (a) [(e)]. }	
\label{Fig2}
\end{figure}


\noindent
\emph{Results---} As an illustrative instance, Fig.~\ref{Fig2}a-d exemplify the two distinctive {\it e-h} pair excitations with the same $\VEC{k}_{ex}$ along the $k_y$-direction, i.e. $\overline{\Gamma_{ex} M_{ex,2}}$, which are excited from the valence $K$ to the conduction $Q'_{2}$ valleys (red arrow line) and from the valence $K'$ to the conduction $Q_{2}$ valleys (blue arrow line), respectively. The two free excitations are presented in the DFT-calculated quasi-particle band structures of Fig.~\ref{Fig2} c and d, and their transition energies are identified to be the same.
For comparative illustration, we consider another set of two inter-band transitions excited from the distinctive valence valleys with the common $\VEC{k}_{ex}$ along $\overline{\Gamma_{ex} K_{ex}'}$, as shown in Fig.~\ref{Fig2}e-h. With the misaligned $\VEC{k}_{ex}$ from $\overline{\Gamma_{ex} M_{ex,i}}$  the transition energies of the two {\it e-h} pair excitations turn out to be different as predicted by the symmetry analysis and identified in Figure~\ref{Fig2}g and h.

\begin{figure}[t]
\includegraphics[width=0.9\columnwidth]{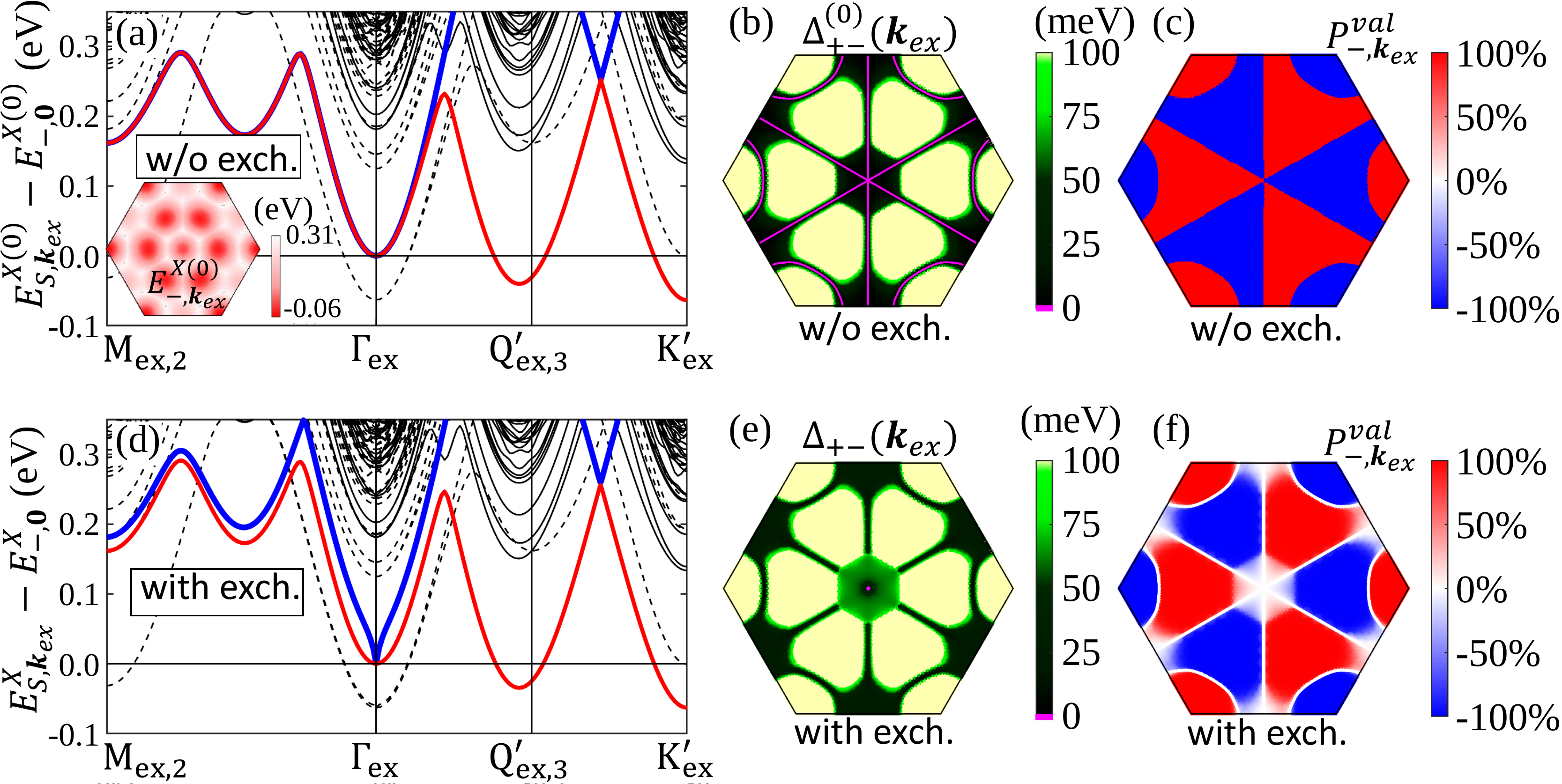}
\caption{(a) Exciton band structure calculated with the neglect of the {\it e-h} exchange interaction along the paths of $\overline{\Gamma _{ex} M _{ex,2}}$ and $\overline{\Gamma_{ex} Q_{ex,3}^{'} K ^{\prime}_{ex}}$. Solid (dashed) lines: spin-like (-unlike) exciton bands.  Red (blue) curves: the lower (upper) band $|-, \VEC{k}_{ex}\rangle$ ($|+, \VEC{k}_{ex}\rangle$) of the lowest spin-like exciton doublet. Inset: the full-zone contour plot of the lowest spin-like exciton band. (b) Contour plot of the energy differences between the lowest spin-like exciton doublet, $\Delta_{+-}(\VEC{k}_{ex}) \equiv E_{+,\VEC{k}_{ex}}^{X} - E_{-,\VEC{k}_{ex}}^{X}$, over the BZ in the exciton-momentum space. The magenta-coloured line indicates the vanishing splitting ($\Delta_{+-}(\VEC{k}_{ex})=0$).  (c) Contour plot of the valley polarization, $P_{-,\VEC{k}_{ex}}^{val}$, of the lowest spin-like exciton states over the BZ. (d)-(f) same as the graphs of (a)-(c) but under the {\it full} consideration of both direct and exchange Coulomb interaction. 
}	
 \label{Fig3}
\end{figure}

Beyond the non-interacting {\it e-h} pair states, the symmetry analysis above remains valid for the exchange-free exciton states. Figure \ref{Fig3}a shows the calculated energy band dispersions of exchange-free exciton, $E_{S,\VEC{k}_{ex}}^{X(0)}$, with the $\VEC{k}_{ex}$ along $\overline{\Gamma _{ex}M _{ex,2}}$ and $\overline{\Gamma _{ex}K_{ex}'}$ directions, solved from the exchange-free BSE including the direct part of Coulomb interaction only. In the absence of {\it e-h} exchange interaction, the energy bands of the lowest exciton doublet along $\overline{\Gamma _{ex}M _{ex,2}}$ does remain degenerate while the ones along $\overline{\Gamma _{ex}K_{ex}'}$ are shown valley-split in energy.
Figure~\ref{Fig3}b presents the energy splitting of the lowest exchange-free spin-like exciton doublet ($S=+/-$ stands for the upper/lower band), $\Delta_{+-}(\VEC{k}_{ex}) \equiv E_{+,\VEC{k}_{ex}}^{X} - E_{-,\VEC{k}_{ex}}^{X}$, as a function of $\VEC{k}_{ex}$ over the BZ, indeed showing the vanishing splitting (indicated by magenta lines) of the exchange-free exciton states lying at the three $\overline{\Gamma _{ex}M _{ex,i}}$ axes.

Figure~\ref{Fig3}d and e shows the calculated exciton bands and the energy difference between the lowest spin-like doublet of WSe$_2$-ML, respectively, with the {\it full} consideration of the both {\it e-h} direct and exchange interactions. Under the influence of {\it e-h} exchange interaction, the exciton band along the $\overline{\Gamma _{ex}M _{ex,i}}$ path is no longer degenerate.
Writing a valley-mixed exciton state as $|S,\VEC{k}_{ex}\rangle = \sum_{\tau=K,K'}\alpha_{S , \VEC{k}_{ex}}^{\tau}|\tau, \VEC{k}_{ex}\rangle $, a linear combination of the exchange-free states with the well-defined valley character, $\{ |\tau,\VEC{k}_{ex}\rangle \}$, the degree of valley polarization of the state is measured by $P_{S, \VEC{k}_{ex}}^{val}\equiv \frac{|\alpha_{S , \VEC{k}_{ex}}^{K}|^2-|\alpha_{S , \VEC{k}_{ex}}^{K'}|^2}{|\alpha_{S , \VEC{k}_{ex}}^{K}|^2 + |\alpha_{S , \VEC{k}_{ex}}^{K'}|^2} $.
Accompanied by the exchange-induced splittings, the exciton states lying on the $\overline{\Gamma _{ex}M _{ex,i}}$ path become highly valley-mixed and featured by $P_{-, \VEC{k}_{ex}}^{val}\sim 0$, as one can see Fig.~\ref{Fig3}f in comparison with Fig.~\ref{Fig3}c.  By contrast, the exciton states lying apart from the $\overline{\Gamma _{ex}M _{ex,i}}$ paths such as those around $Q_{ex,i}/Q_{ex,i}'$ and $K_{ex}/K_{ex}'$ valleys yet well maintain the superior high degrees of the valley polarizations with $\mid P_{-, \VEC{k}_{ex}}^{val} \mid \lesssim 100\% $.

\begin{figure}[t]
\includegraphics[width=0.9\columnwidth]{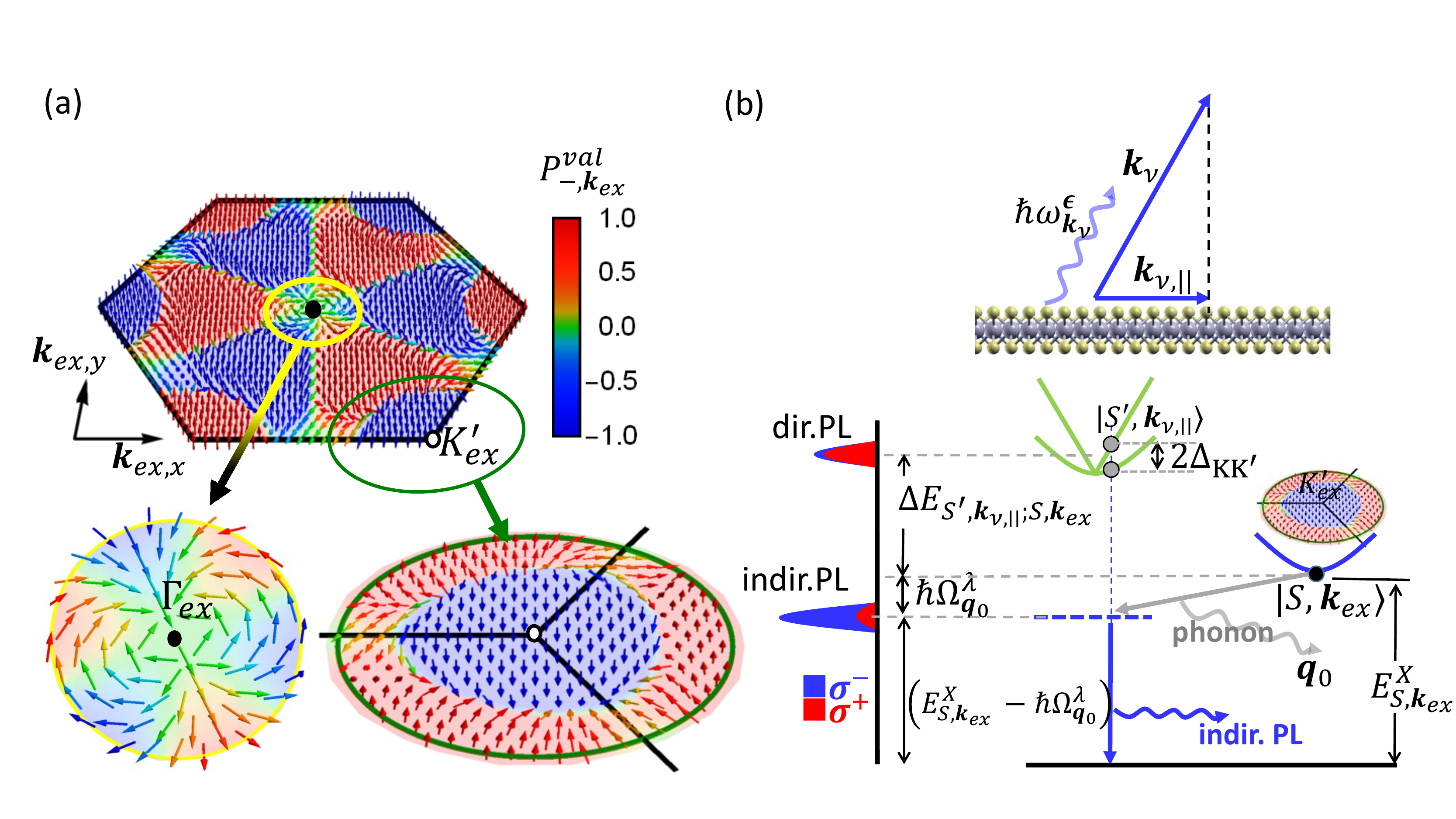}
\caption{(a) Valley-pseudospin texture of the valley-polarized exciton states over the entire BZ. Left (Right) inset: Zoom-in view of the pseudospin texture around the $\Gamma_{ex}$  ($K_{ex}^{\prime}$) point. (b) Schematics of a phonon-assisted indirect PL process. The upper inset depicts the non-vertically emitted light with $\VEC{k}_{\nu,\mid\mid} \neq \VEC{0}$ from an indirect PL, in coincidence with the wave vector of the intermediate bright states $|S^{\prime}, \VEC{k}_{ex} = \VEC{k}_{\nu,\mid\mid} \rangle$.}	
\label{Fig4}
\end{figure}

To further recognize the quantum nature of the valley-polarized exciton states, we reformulate the numerically calculated lowest exciton states (with the band index of $S=-$) as a Bloch vector, $\left| S, \VEC{k} _{ex} \right\rangle = - e ^{- i \phi _{S , \VEC{k} _{ex}}} \sin \frac{\theta _{S , \VEC{k} _{ex}}}{2} \left| K , \VEC{k} _{ex} \right\rangle + \cos \frac{\theta _{S , \VEC{k} _{ex}}}{2} \left| K ^{\prime} , \VEC{k} _{ex} \right\rangle$, in terms of the geometrical angles $\theta_{S , \VEC{k}_{ex}}$ and $\phi_{S , \VEC{k}_{ex}}$, allowing for visualization in the Bloch sphere (see Supplementary Materials \cite{Supplement} for details). Figure \ref{Fig4}a presents the Bloch vectors of the lowest spin-like exciton states over the BZ forming a valley-pseudospin texture.
Around the $\Gamma_{ex}$-point, the phase angle $\phi_{S , \VEC{k}_{ex}}$ is evolved with a winding number $n_w=2$ by the in-plane rotation of $\VEC{k}_{ex}$, reflecting the dipole-dipole interacting nature of the {\it e-h} exchange interaction. \cite{WYao2014b, SGLouie2015a} Governed by the symmetries of $D_{3h}$, the valley pseudo-spin texture of the full-zone exciton band in the momentum space is anisotropically patterned by individual skyrmion-like structures centred with highly valley-polarized exciton states in the $Q_{ex,i}/Q_{ex,i}'$ and $K_{ex}/K_{ex}'$ excitonic valleys as seen in Figure \ref{Fig4}a.
In spite of the violation of the momentum selection rules, those inter-valley exciton states inherited with the high degree of valley polarization could emit light via the two-step transition processes with the involvement of phonon- or defect-scattering.
Experimentally, the indirect PLs from the lowest inter-valley exciton states with $\VEC{k}_{ex} \in K_{ex}/K_{ex}'$ have been recently observed and, interestingly, present superiorly high optical polarization. \cite{SFShi2019, CHLui2020, WYao2020a}

Considering the phonon and photon reservoirs and their couplings to excitons,\cite{EMalic2020a}
the total Hamiltonian of the extended exciton-photon-phonon system reads $H=H_X + H_{\nu} + H_{ph}+ H_{X-\nu} + H_{X-ph}$, where $H_X=\sum_{S\VEC{k}_{ex}} E_{S,\VEC{k}_{ex}}^X \hat{X}_{S,\VEC{k}_{ex}}^{\dagger} \hat{X}_{S,\VEC{k}_{ex}}$ stands for the single-exciton Hamiltonian, $\hat{X}$ ($\hat{X}^{\dagger}$) is the operator annihilating (creating) an exciton, $H_{\nu}=\sum_{\VEC{\epsilon}}\sum_{\VEC{k}_{\nu}} \hbar \omega_{ \VEC{k}_{\nu}}^{\VEC{\epsilon}} a_{\VEC{\epsilon}, \VEC{k}_{\nu}}^\dagger a_{\VEC{\epsilon}, \VEC{k}_{\nu}}$ ($H_{ph}=\sum_{\lambda}\sum_{\VEC{q}} \hbar \Omega_{\VEC{q}}^{\lambda} b_{\lambda,\VEC{q}}^\dagger b_{\lambda,\VEC{q}}$) is the Hamiltonian of photon (phonon) reservoir, $\omega_{\VEC{k}_{\nu}}^{\VEC{\epsilon}}$ is the frequency of the $\VEC{\epsilon}$-polarized photon with the wave vector $\VEC{k}_{\nu}$ and $a$ ($a^\dagger$) is the particle operator that annihilates (creates) the photon, $\Omega_{\VEC{q}}^{\lambda}$ is the frequency of the $\lambda$-kind of phonon with the wave vector $\VEC{q}$,\cite{jin2014intrinsic} and $b$ ($b^\dagger$) is the particle operator that annihilates (creates) the phonon.
$H_{X-\nu}$ ($H_{X-ph}$) is the Hamiltonian of exciton-photon (exciton-phonon) coupling in terms of the coupling constants $\eta_{S,\VEC{k}_{ex}}$ ($g_{S',\VEC{k}_{ex}';S,\VEC{k}_{ex}}$) that converts an exciton in the bright state $\mid S,\VEC{k}_{ex}\rangle $ to a photon (that couples the exciton states $|S,\VEC{k}_{ex}\rangle$ and $|S',\VEC{k}_{ex}'\rangle$). The complete formalisms of $H_{X-\nu}$ and $H_{X-ph}$ are explicitly given in Supplementary Material. \cite{Supplement}

From the second-order perturbation theory, the transition rate of the phonon-assisted PL is evaluated by
\begin{equation}
\Gamma _{f \leftarrow i} ^{(2)} = \frac{2 \pi}{\hbar} \left| \sum _{m} \frac{\left\langle f \right| \op{H} _{X-\nu} \left| m \right\rangle \left\langle m \right| \op{H} _{X-ph} \left| i \right\rangle}{E_{i} - E_{m} + i 0 ^{+}} \right| ^{2} \delta \left( E _{f} - E _{i} \right) \, ,
\end{equation}
where $|i\rangle$ denotes the initial state, $|m\rangle$ the intermediate states following the emission or absorption of a phonon, and $|f\rangle$ is the final state of the indirect PL.
Specifically, we shall analyze the transition rate and polarization of the indirect PL emitting the photons with a given wave vector $\VEC{k}_{\nu}$ from some exciton initial state, $|S, \VEC{k}_{ex}\rangle$. In the situation, the wave vectors of the intermediate bright exciton states, $|S'=\pm,\VEC{k}_{ex}'\rangle$, and the phonons really involved in the transition process are deterministic under the law of momentum conservation, i.e. $\VEC{k}_{ex}'=\VEC{k}_{\nu,\mid\mid}$ and $\VEC{q}=\VEC{k}_{ex} - \VEC{k}_{\nu,\mid\mid} \equiv \VEC{q}_{0}$, as illustrated in Fig.~\ref{Fig4}b.

Considering a valley-mixed initial state, $|S,\VEC{k}_{ex} \rangle= \alpha_{S,\VEC{k}_{ex}}^{K}|K,\VEC{k}_{ex}\rangle + \alpha_{S,\VEC{k}_{ex}}^{K'}|K',\VEC{k}_{ex}\rangle$, the transition rates of the $\sigma^+$- and $\sigma^-$-polarized indirect PL's via the intermediate bright exciton doublet split by  $\Delta_{+-}(\VEC{k}_{\nu,||})=2|\widetilde{\Delta}_{K K'}(\VEC{k}_{\nu,||})| $ are derived as
$\Gamma _{S, \VEC{k} _{ex}} ^{(2)} \!\left( \sigma ^{\pm} \right) \approx \, \gamma _{0} ^{(2)} \left| \alpha _{S, \VEC{k} _{ex}} ^{K _{\pm}} + \widetilde{\beta} _{\VEC{k} _{\nu , \parallel}} ^{K _{\pm} K _{\mp}} \alpha_{S,\VEC{k}_{ex}}^{K_{\mp}} \right| ^{2} \times \delta \left( E _{S, \VEC{k} _{ex}} ^{X} - \hbar \Omega _{\VEC{q} _{0}} ^{\lambda} - \hbar \omega _{\VEC{k} _{\nu}}^{\VEC{\epsilon}} \right) $,
where the symbol $K_{+}$ ($K_{-}$) are introduced to denote the $K$- ($K'$)-valley,  $\gamma_0^{(2)}$ is the averaged transition rate of the polarization-unresolved indirect PL,  $\widetilde{\beta} _{\VEC{k} _{\nu , \parallel}} ^{K _{\pm} K _{\mp}} \approx - \frac{\widetilde{\Delta}_{K_{\pm} K_{\mp}} \left( \VEC{k} _{\nu , \parallel} \right)}{\Delta E_{-, \VEC{k}_{\nu , \parallel}; S, \VEC{k}_{ex}} + \hbar \Omega_{\VEC{q} _0} ^{\lambda}}$,
$\widetilde{\Delta}_{K_{\pm} K_{\mp}}$ is the matrix element of {\it e-h} exchange interaction that couples the $K_{\pm}$- and $K_{\mp}$-valley exciton and $\Delta E_{-, \VEC{k}_{\nu , \parallel}; S, \VEC{k}_{ex}} \equiv E _{-, \VEC{k} _{\nu , \parallel}} ^{X} - E _{S, \VEC{k} _{ex}} ^{X} $.\cite{Supplement}
 Accordingly, one can show that the optical polarization, defined by $P _{S, \VEC{k} _{ex}}^{op(2)} \equiv \frac{\Gamma _{S, \VEC{k} _{ex}} ^{(2)} \left(\sigma ^{+} \right) - \Gamma _{S, \VEC{k} _{ex}} ^{(2)} \left( \sigma ^{-} \right)}{\Gamma _{S, \VEC{k} _{ex}} ^{(2)} \left( \sigma ^{+} \right) + \Gamma _{S, \VEC{k} _{ex}} ^{(2)} \left( \sigma ^{-} \right)} $, of the indirect PL from $|S,\VEC{k}_{ex} \rangle$ is given by \cite{Supplement}

\begin{equation}
P _{S, \VEC{k} _{ex}}^{op(2)}  \approx P _{S, \VEC{k} _{ex}} ^{val} \left( 1 - 4 \Re \left[ \widetilde{\beta}_{\VEC{k}_{\nu, \parallel}}^{KK'} \left( \alpha _{S, \VEC{k} _{ex}} ^{K} \right) ^{*} \alpha _{S, \VEC{k} _{ex}} ^{K ^{\prime}} \right] \right)\, ,
\label{Pconvert}
\end{equation}
which accounts for the degree of conversion of the valley-polarization of an inter-valley exciton to the optical polarization of the resulting indirect PL. For WSe$_2$-ML's, $|\widetilde{\Delta}_{K K'}(\VEC{k}_{\nu,||})| \sim 1$meV \cite{SGLouie2015a} and $\left( \Delta E_{-, \VEC{k}_{\nu , \parallel}; S, \VEC{k}_{ex}} + \hbar \Omega _{\VEC{q} _{0}} ^{\lambda}\right) \sim 60$meV.\cite{CHLui2020, SFShi2019, chen2020}  The latter is measurable from the energy difference between the direct and indirect PLs from bright exciton and inter-valley one, respectively, as illustrated by Fig.~\ref{Fig4}b. \cite{CHLui2020, SFShi2019, chen2020} Accordingly, one can estimate $|\widetilde{\beta} _{\VEC{k} _{\nu , \parallel}}^{KK'}| \ll 0.1$, with which the optical polarization of the indirect PL is shown nearly the same as the degree of valley polarization of the initial inter-valley exciton state according to Eq.~\ref{Pconvert}. Note that the effect of severe valley-depolarization in the intermediate bright states is found much suppressed in the second-order PL process.
As a result, the valley-polarization of the inter-valley exciton state is actually excellently well transferable to the resulting optical polarization. This might account for the recently observed highly polarized indirect PL from tungsten-based TMD-MLs. \cite{WYao2020a, CHLui2020, SFShi2019}


In conclusion, we present a theoretical and computational investigation of the complete full-zone exciton band structures of TMD-MLs.
While the bright exciton states of TMD-MLs lying around the central of the BZ are known to be inherently valley-depolarized, our studies reveal that most finite-momentum exciton states over the BZ are yet well immune from the exchange-induced valley depolarization, except for those with specific exciton momenta directionally in coincidence with the $C_2'$- and $\sigma_v$-associated axes. Governed by the symmetries, the valley pseudo-spin texture of the exciton states over the entire BZ is locally featured by skyrmion-like structures surrounding the highly valley-polarized inter-valley exciton states. Importantly, the superior valley polarizations of those inter-valley excitons are shown excellently transferable to the optical polarization under the assistance of phonons. The finding sheds light on the prospective of the valley-based photonics with the utilization of those inter-valley finite momentum excitons.

P.Y.L. and S.J.C. thank M. Bieniek and P. Hawrylak for fruitful discussion.
This study is supported by the Ministry of Science and Technology, Taiwan, under contracts, MOST 109-2639-E-009-001 and 109-2112-M-009 -018 -MY3, and by National Center for High-Performance Computing (NCHC), Taiwan.

\bibliography{full-zone-exciton-band-refs}{}

\begin{thebibliography}{50}
\expandafter\ifx\csname natexlab\endcsname\relax\def\natexlab#1{#1}\fi
\expandafter\ifx\csname bibnamefont\endcsname\relax
  \def\bibnamefont#1{#1}\fi
\expandafter\ifx\csname bibfnamefont\endcsname\relax
  \def\bibfnamefont#1{#1}\fi
\expandafter\ifx\csname citenamefont\endcsname\relax
  \def\citenamefont#1{#1}\fi
\expandafter\ifx\csname url\endcsname\relax
  \def\url#1{\texttt{#1}}\fi
\expandafter\ifx\csname urlprefix\endcsname\relax\def\urlprefix{URL }\fi
\providecommand{\bibinfo}[2]{#2}
\providecommand{\eprint}[2][]{\url{#2}}

\bibitem[{\citenamefont{Mak et~al.}(2010)\citenamefont{Mak, Lee, Hone, Shan,
  and Heinz}}]{TFHeinz2010}
\bibinfo{author}{\bibfnamefont{K.~F.} \bibnamefont{Mak}},
  \bibinfo{author}{\bibfnamefont{C.}~\bibnamefont{Lee}},
  \bibinfo{author}{\bibfnamefont{J.}~\bibnamefont{Hone}},
  \bibinfo{author}{\bibfnamefont{J.}~\bibnamefont{Shan}}, \bibnamefont{and}
  \bibinfo{author}{\bibfnamefont{T.~F.} \bibnamefont{Heinz}},
  \bibinfo{journal}{Phys. Rev. Lett.} \textbf{\bibinfo{volume}{105}},
  \bibinfo{pages}{136805} (\bibinfo{year}{2010}),
  \urlprefix\url{https://link.aps.org/doi/10.1103/PhysRevLett.105.136805}.

\bibitem[{\citenamefont{Xiao et~al.}(2012)\citenamefont{Xiao, Liu, Feng, Xu,
  and Yao}}]{WYao2012a}
\bibinfo{author}{\bibfnamefont{D.}~\bibnamefont{Xiao}},
  \bibinfo{author}{\bibfnamefont{G.-B.} \bibnamefont{Liu}},
  \bibinfo{author}{\bibfnamefont{W.}~\bibnamefont{Feng}},
  \bibinfo{author}{\bibfnamefont{X.}~\bibnamefont{Xu}}, \bibnamefont{and}
  \bibinfo{author}{\bibfnamefont{W.}~\bibnamefont{Yao}},
  \bibinfo{journal}{Phys. Rev. Lett.} \textbf{\bibinfo{volume}{108}},
  \bibinfo{pages}{196802} (\bibinfo{year}{2012}),
  \urlprefix\url{https://link.aps.org/doi/10.1103/PhysRevLett.108.196802}.

\bibitem[{\citenamefont{Xu et~al.}(2014)\citenamefont{Xu, Yao, Xiao, and
  Heinz}}]{WYao2014a}
\bibinfo{author}{\bibfnamefont{X.}~\bibnamefont{Xu}},
  \bibinfo{author}{\bibfnamefont{W.}~\bibnamefont{Yao}},
  \bibinfo{author}{\bibfnamefont{D.}~\bibnamefont{Xiao}}, \bibnamefont{and}
  \bibinfo{author}{\bibfnamefont{T.~F.} \bibnamefont{Heinz}},
  \bibinfo{journal}{Nat. Phys.} \textbf{\bibinfo{volume}{10}},
  \bibinfo{pages}{343} (\bibinfo{year}{2014}),
  \urlprefix\url{https://doi.org/10.1038/nphys2942}.

\bibitem[{\citenamefont{Yu et~al.}(2014)\citenamefont{Yu, Liu, Gong, Xu, and
  Yao}}]{WYao2014b}
\bibinfo{author}{\bibfnamefont{H.}~\bibnamefont{Yu}},
  \bibinfo{author}{\bibfnamefont{G.-B.} \bibnamefont{Liu}},
  \bibinfo{author}{\bibfnamefont{P.}~\bibnamefont{Gong}},
  \bibinfo{author}{\bibfnamefont{X.}~\bibnamefont{Xu}}, \bibnamefont{and}
  \bibinfo{author}{\bibfnamefont{W.}~\bibnamefont{Yao}}, \bibinfo{journal}{Nat.
  Commun.} \textbf{\bibinfo{volume}{5}}, \bibinfo{pages}{3876}
  (\bibinfo{year}{2014}), \urlprefix\url{https://doi.org/10.1038/ncomms4876}.

\bibitem[{\citenamefont{Zeng et~al.}(2012)\citenamefont{Zeng, Dai, Yao, Xiao,
  and Cui}}]{WYao2012b}
\bibinfo{author}{\bibfnamefont{H.}~\bibnamefont{Zeng}},
  \bibinfo{author}{\bibfnamefont{J.}~\bibnamefont{Dai}},
  \bibinfo{author}{\bibfnamefont{W.}~\bibnamefont{Yao}},
  \bibinfo{author}{\bibfnamefont{D.}~\bibnamefont{Xiao}}, \bibnamefont{and}
  \bibinfo{author}{\bibfnamefont{X.}~\bibnamefont{Cui}}, \bibinfo{journal}{Nat.
  Nanotech.} \textbf{\bibinfo{volume}{7}}, \bibinfo{pages}{490}
  (\bibinfo{year}{2012}),
  \urlprefix\url{https://doi.org/10.1038/nnano.2012.95}.

\bibitem[{\citenamefont{Mak et~al.}(2018)\citenamefont{Mak, Xiao, and
  Shan}}]{JShan2018a}
\bibinfo{author}{\bibfnamefont{K.~F.} \bibnamefont{Mak}},
  \bibinfo{author}{\bibfnamefont{D.}~\bibnamefont{Xiao}}, \bibnamefont{and}
  \bibinfo{author}{\bibfnamefont{J.}~\bibnamefont{Shan}},
  \bibinfo{journal}{Nat. Photon.} \textbf{\bibinfo{volume}{12}},
  \bibinfo{pages}{451} (\bibinfo{year}{2018}),
  \urlprefix\url{https://doi.org/10.1038/s41566-018-0204-6}.

\bibitem[{\citenamefont{Wang et~al.}(2018)\citenamefont{Wang, Chernikov,
  Glazov, Heinz, Marie, Amand, and Urbaszek}}]{GWang2015}
\bibinfo{author}{\bibfnamefont{G.}~\bibnamefont{Wang}},
  \bibinfo{author}{\bibfnamefont{A.}~\bibnamefont{Chernikov}},
  \bibinfo{author}{\bibfnamefont{M.~M.} \bibnamefont{Glazov}},
  \bibinfo{author}{\bibfnamefont{T.~F.} \bibnamefont{Heinz}},
  \bibinfo{author}{\bibfnamefont{X.}~\bibnamefont{Marie}},
  \bibinfo{author}{\bibfnamefont{T.}~\bibnamefont{Amand}}, \bibnamefont{and}
  \bibinfo{author}{\bibfnamefont{B.}~\bibnamefont{Urbaszek}},
  \bibinfo{journal}{Rev. Mod. Phys.} \textbf{\bibinfo{volume}{90}},
  \bibinfo{pages}{021001} (\bibinfo{year}{2018}),
  \urlprefix\url{https://doi.org/10.1103/RevModPhys.90.021001}.

\bibitem[{\citenamefont{Smole\'nski et~al.}(2016)\citenamefont{Smole\'nski,
  Goryca, Koperski, Faugeras, Kazimierczuk, Bogucki, Nogajewski, Kossacki, and
  Potemski}}]{Smolenski2016}
\bibinfo{author}{\bibfnamefont{T.}~\bibnamefont{Smole\'nski}},
  \bibinfo{author}{\bibfnamefont{M.}~\bibnamefont{Goryca}},
  \bibinfo{author}{\bibfnamefont{M.}~\bibnamefont{Koperski}},
  \bibinfo{author}{\bibfnamefont{C.}~\bibnamefont{Faugeras}},
  \bibinfo{author}{\bibfnamefont{T.}~\bibnamefont{Kazimierczuk}},
  \bibinfo{author}{\bibfnamefont{A.}~\bibnamefont{Bogucki}},
  \bibinfo{author}{\bibfnamefont{K.}~\bibnamefont{Nogajewski}},
  \bibinfo{author}{\bibfnamefont{P.}~\bibnamefont{Kossacki}}, \bibnamefont{and}
  \bibinfo{author}{\bibfnamefont{M.}~\bibnamefont{Potemski}},
  \bibinfo{journal}{Phys. Rev. X} \textbf{\bibinfo{volume}{6}},
  \bibinfo{pages}{021024} (\bibinfo{year}{2016}),
  \urlprefix\url{https://doi.org/10.1103/PhysRevX.6.021024}.

\bibitem[{\citenamefont{Yu and Wu}(2014)}]{MWWu2014}
\bibinfo{author}{\bibfnamefont{T.}~\bibnamefont{Yu}} \bibnamefont{and}
  \bibinfo{author}{\bibfnamefont{M.~W.} \bibnamefont{Wu}},
  \bibinfo{journal}{Phys. Rev. B} \textbf{\bibinfo{volume}{89}},
  \bibinfo{pages}{205303} (\bibinfo{year}{2014}),
  \urlprefix\url{https://link.aps.org/doi/10.1103/PhysRevB.89.205303}.

\bibitem[{\citenamefont{Hao et~al.}(2016)\citenamefont{Hao, Moody, Wu, Dass,
  Xu, Chen, Sun, Li, Li, MacDonald et~al.}}]{AHMacDonald2016}
\bibinfo{author}{\bibfnamefont{K.}~\bibnamefont{Hao}},
  \bibinfo{author}{\bibfnamefont{G.}~\bibnamefont{Moody}},
  \bibinfo{author}{\bibfnamefont{F.}~\bibnamefont{Wu}},
  \bibinfo{author}{\bibfnamefont{C.~K.} \bibnamefont{Dass}},
  \bibinfo{author}{\bibfnamefont{L.}~\bibnamefont{Xu}},
  \bibinfo{author}{\bibfnamefont{C.-H.} \bibnamefont{Chen}},
  \bibinfo{author}{\bibfnamefont{L.}~\bibnamefont{Sun}},
  \bibinfo{author}{\bibfnamefont{M.-Y.} \bibnamefont{Li}},
  \bibinfo{author}{\bibfnamefont{L.-J.} \bibnamefont{Li}},
  \bibinfo{author}{\bibfnamefont{A.~H.} \bibnamefont{MacDonald}},
  \bibnamefont{et~al.}, \bibinfo{journal}{Nat. Phys.}
  \textbf{\bibinfo{volume}{12}}, \bibinfo{pages}{677} (\bibinfo{year}{2016}),
  \urlprefix\url{https://doi.org/10.1038/nphys3674}.

\bibitem[{\citenamefont{Wang et~al.}(2014)\citenamefont{Wang, Bouet, Lagarde,
  Vidal, Balocchi, Amand, Marie, and Urbaszek}}]{GWang2014}
\bibinfo{author}{\bibfnamefont{G.}~\bibnamefont{Wang}},
  \bibinfo{author}{\bibfnamefont{L.}~\bibnamefont{Bouet}},
  \bibinfo{author}{\bibfnamefont{D.}~\bibnamefont{Lagarde}},
  \bibinfo{author}{\bibfnamefont{M.}~\bibnamefont{Vidal}},
  \bibinfo{author}{\bibfnamefont{A.}~\bibnamefont{Balocchi}},
  \bibinfo{author}{\bibfnamefont{T.}~\bibnamefont{Amand}},
  \bibinfo{author}{\bibfnamefont{X.}~\bibnamefont{Marie}}, \bibnamefont{and}
  \bibinfo{author}{\bibfnamefont{B.}~\bibnamefont{Urbaszek}},
  \bibinfo{journal}{Phys. Rev. B} \textbf{\bibinfo{volume}{90}},
  \bibinfo{pages}{075413} (\bibinfo{year}{2014}),
  \urlprefix\url{https://doi.org/10.1103/PhysRevB.90.075413}.

\bibitem[{\citenamefont{Glazov et~al.}(2015)\citenamefont{Glazov, Ivchenko,
  Wang, Amand, Marie, Urbaszek, and Liu}}]{Glazov2015}
\bibinfo{author}{\bibfnamefont{M.}~\bibnamefont{Glazov}},
  \bibinfo{author}{\bibfnamefont{E.}~\bibnamefont{Ivchenko}},
  \bibinfo{author}{\bibfnamefont{G.}~\bibnamefont{Wang}},
  \bibinfo{author}{\bibfnamefont{T.}~\bibnamefont{Amand}},
  \bibinfo{author}{\bibfnamefont{X.}~\bibnamefont{Marie}},
  \bibinfo{author}{\bibfnamefont{B.}~\bibnamefont{Urbaszek}}, \bibnamefont{and}
  \bibinfo{author}{\bibfnamefont{B.}~\bibnamefont{Liu}},
  \bibinfo{journal}{Phys. Status Solidi (b)} \textbf{\bibinfo{volume}{252}},
  \bibinfo{pages}{2349} (\bibinfo{year}{2015}).

\bibitem[{\citenamefont{Selig et~al.}(2020)\citenamefont{Selig, Katsch, Brem,
  Mkrtchian, Malic, and Knorr}}]{selig2020suppression}
\bibinfo{author}{\bibfnamefont{M.}~\bibnamefont{Selig}},
  \bibinfo{author}{\bibfnamefont{F.}~\bibnamefont{Katsch}},
  \bibinfo{author}{\bibfnamefont{S.}~\bibnamefont{Brem}},
  \bibinfo{author}{\bibfnamefont{G.~F.} \bibnamefont{Mkrtchian}},
  \bibinfo{author}{\bibfnamefont{E.}~\bibnamefont{Malic}}, \bibnamefont{and}
  \bibinfo{author}{\bibfnamefont{A.}~\bibnamefont{Knorr}},
  \bibinfo{journal}{Physical Review Research} \textbf{\bibinfo{volume}{2}},
  \bibinfo{pages}{023322} (\bibinfo{year}{2020}).

\bibitem[{\citenamefont{Qiu et~al.}(2015)\citenamefont{Qiu, Cao, and
  Louie}}]{SGLouie2015a}
\bibinfo{author}{\bibfnamefont{D.~Y.} \bibnamefont{Qiu}},
  \bibinfo{author}{\bibfnamefont{T.}~\bibnamefont{Cao}}, \bibnamefont{and}
  \bibinfo{author}{\bibfnamefont{S.~G.} \bibnamefont{Louie}},
  \bibinfo{journal}{Phys. Rev. Lett.} \textbf{\bibinfo{volume}{115}},
  \bibinfo{pages}{176801} (\bibinfo{year}{2015}),
  \urlprefix\url{https://link.aps.org/doi/10.1103/PhysRevLett.115.176801}.

\bibitem[{\citenamefont{Li et~al.}(2019)\citenamefont{Li, Wang, Jin, Lu, Lian,
  Meng, Blei, Gao, Taniguchi, Watanabe et~al.}}]{SFShi2019}
\bibinfo{author}{\bibfnamefont{Z.}~\bibnamefont{Li}},
  \bibinfo{author}{\bibfnamefont{T.}~\bibnamefont{Wang}},
  \bibinfo{author}{\bibfnamefont{C.}~\bibnamefont{Jin}},
  \bibinfo{author}{\bibfnamefont{Z.}~\bibnamefont{Lu}},
  \bibinfo{author}{\bibfnamefont{Z.}~\bibnamefont{Lian}},
  \bibinfo{author}{\bibfnamefont{Y.}~\bibnamefont{Meng}},
  \bibinfo{author}{\bibfnamefont{M.}~\bibnamefont{Blei}},
  \bibinfo{author}{\bibfnamefont{M.}~\bibnamefont{Gao}},
  \bibinfo{author}{\bibfnamefont{T.}~\bibnamefont{Taniguchi}},
  \bibinfo{author}{\bibfnamefont{K.}~\bibnamefont{Watanabe}},
  \bibnamefont{et~al.}, \bibinfo{journal}{ACS Nano}
  \textbf{\bibinfo{volume}{13}}, \bibinfo{pages}{14107} (\bibinfo{year}{2019}),
  \urlprefix\url{https://doi.org/10.1021/acsnano.9b06682}.

\bibitem[{\citenamefont{Liu et~al.}(2020)\citenamefont{Liu, van Baren, Liang,
  Taniguchi, Watanabe, Gabor, Chang, and Lui}}]{CHLui2020}
\bibinfo{author}{\bibfnamefont{E.}~\bibnamefont{Liu}},
  \bibinfo{author}{\bibfnamefont{J.}~\bibnamefont{van Baren}},
  \bibinfo{author}{\bibfnamefont{C.-T.} \bibnamefont{Liang}},
  \bibinfo{author}{\bibfnamefont{T.}~\bibnamefont{Taniguchi}},
  \bibinfo{author}{\bibfnamefont{K.}~\bibnamefont{Watanabe}},
  \bibinfo{author}{\bibfnamefont{N.~M.} \bibnamefont{Gabor}},
  \bibinfo{author}{\bibfnamefont{Y.-C.} \bibnamefont{Chang}}, \bibnamefont{and}
  \bibinfo{author}{\bibfnamefont{C.~H.} \bibnamefont{Lui}},
  \bibinfo{journal}{Phys. Rev. Lett.} \textbf{\bibinfo{volume}{124}},
  \bibinfo{pages}{196802} (\bibinfo{year}{2020}),
  \urlprefix\url{https://link.aps.org/doi/10.1103/PhysRevLett.124.196802}.

\bibitem[{\citenamefont{He et~al.}(2020)\citenamefont{He, Rivera, Tuan, Wilson,
  Yang, Taniguchi, Watanabe, Yan, Mandrus, Yu et~al.}}]{WYao2020a}
\bibinfo{author}{\bibfnamefont{M.}~\bibnamefont{He}},
  \bibinfo{author}{\bibfnamefont{P.}~\bibnamefont{Rivera}},
  \bibinfo{author}{\bibfnamefont{D.~V.} \bibnamefont{Tuan}},
  \bibinfo{author}{\bibfnamefont{N.~P.} \bibnamefont{Wilson}},
  \bibinfo{author}{\bibfnamefont{M.}~\bibnamefont{Yang}},
  \bibinfo{author}{\bibfnamefont{T.}~\bibnamefont{Taniguchi}},
  \bibinfo{author}{\bibfnamefont{K.}~\bibnamefont{Watanabe}},
  \bibinfo{author}{\bibfnamefont{J.}~\bibnamefont{Yan}},
  \bibinfo{author}{\bibfnamefont{D.~G.} \bibnamefont{Mandrus}},
  \bibinfo{author}{\bibfnamefont{H.}~\bibnamefont{Yu}}, \bibnamefont{et~al.},
  \bibinfo{journal}{Nat. Commun.} \textbf{\bibinfo{volume}{11}},
  \bibinfo{pages}{618} (\bibinfo{year}{2020}),
  \urlprefix\url{https://doi.org/10.1038/s41467-020-14472-0}.

\bibitem[{\citenamefont{Brem et~al.}(2020)\citenamefont{Brem, Ekman,
  Christiansen, Katsch, Selig, Robert, Marie, Urbaszek, Knorr, and
  Malic}}]{EMalic2020a}
\bibinfo{author}{\bibfnamefont{S.}~\bibnamefont{Brem}},
  \bibinfo{author}{\bibfnamefont{A.}~\bibnamefont{Ekman}},
  \bibinfo{author}{\bibfnamefont{D.}~\bibnamefont{Christiansen}},
  \bibinfo{author}{\bibfnamefont{F.}~\bibnamefont{Katsch}},
  \bibinfo{author}{\bibfnamefont{M.}~\bibnamefont{Selig}},
  \bibinfo{author}{\bibfnamefont{C.}~\bibnamefont{Robert}},
  \bibinfo{author}{\bibfnamefont{X.}~\bibnamefont{Marie}},
  \bibinfo{author}{\bibfnamefont{B.}~\bibnamefont{Urbaszek}},
  \bibinfo{author}{\bibfnamefont{A.}~\bibnamefont{Knorr}}, \bibnamefont{and}
  \bibinfo{author}{\bibfnamefont{E.}~\bibnamefont{Malic}},
  \bibinfo{journal}{Nano Letters} \textbf{\bibinfo{volume}{20}},
  \bibinfo{pages}{2849} (\bibinfo{year}{2020}), \bibinfo{note}{pMID: 32084315},
  \urlprefix\url{https://doi.org/10.1021/acs.nanolett.0c00633}.

\bibitem[{\citenamefont{Liu et~al.}(2019)\citenamefont{Liu, van Baren,
  Taniguchi, Watanabe, Chang, and Lui}}]{CHLui2019a}
\bibinfo{author}{\bibfnamefont{E.}~\bibnamefont{Liu}},
  \bibinfo{author}{\bibfnamefont{J.}~\bibnamefont{van Baren}},
  \bibinfo{author}{\bibfnamefont{T.}~\bibnamefont{Taniguchi}},
  \bibinfo{author}{\bibfnamefont{K.}~\bibnamefont{Watanabe}},
  \bibinfo{author}{\bibfnamefont{Y.-C.} \bibnamefont{Chang}}, \bibnamefont{and}
  \bibinfo{author}{\bibfnamefont{C.~H.} \bibnamefont{Lui}},
  \bibinfo{journal}{Phys. Rev. Research} \textbf{\bibinfo{volume}{1}},
  \bibinfo{pages}{032007} (\bibinfo{year}{2019}),
  \urlprefix\url{https://link.aps.org/doi/10.1103/PhysRevResearch.1.032007}.

\bibitem[{\citenamefont{Bao et~al.}(2020)\citenamefont{Bao, del {\'A}guila, Do,
  Liu, Pei, and Xiong}}]{bao2020probing}
\bibinfo{author}{\bibfnamefont{D.}~\bibnamefont{Bao}},
  \bibinfo{author}{\bibfnamefont{A.~G.} \bibnamefont{del {\'A}guila}},
  \bibinfo{author}{\bibfnamefont{T.~T.~H.} \bibnamefont{Do}},
  \bibinfo{author}{\bibfnamefont{S.}~\bibnamefont{Liu}},
  \bibinfo{author}{\bibfnamefont{J.}~\bibnamefont{Pei}}, \bibnamefont{and}
  \bibinfo{author}{\bibfnamefont{Q.}~\bibnamefont{Xiong}}, \bibinfo{journal}{2D
  Materials} \textbf{\bibinfo{volume}{7}}, \bibinfo{pages}{031002}
  (\bibinfo{year}{2020}).

\bibitem[{\citenamefont{Chen et~al.}(2018)\citenamefont{Chen, Goldstein,
  Taniguchi, Watanabe, and Yan}}]{chen2018coulomb}
\bibinfo{author}{\bibfnamefont{S.-Y.} \bibnamefont{Chen}},
  \bibinfo{author}{\bibfnamefont{T.}~\bibnamefont{Goldstein}},
  \bibinfo{author}{\bibfnamefont{T.}~\bibnamefont{Taniguchi}},
  \bibinfo{author}{\bibfnamefont{K.}~\bibnamefont{Watanabe}}, \bibnamefont{and}
  \bibinfo{author}{\bibfnamefont{J.}~\bibnamefont{Yan}}, \bibinfo{journal}{Nat.
  Commun.} \textbf{\bibinfo{volume}{9}}, \bibinfo{pages}{3717}
  (\bibinfo{year}{2018}).

\bibitem[{\citenamefont{Wu et~al.}(2019)\citenamefont{Wu, Chen, Zhou, and
  Zhu}}]{HZhu2019a}
\bibinfo{author}{\bibfnamefont{L.}~\bibnamefont{Wu}},
  \bibinfo{author}{\bibfnamefont{Y.}~\bibnamefont{Chen}},
  \bibinfo{author}{\bibfnamefont{H.}~\bibnamefont{Zhou}}, \bibnamefont{and}
  \bibinfo{author}{\bibfnamefont{H.}~\bibnamefont{Zhu}}, \bibinfo{journal}{ACS
  Nano} \textbf{\bibinfo{volume}{13}}, \bibinfo{pages}{2341}
  (\bibinfo{year}{2019}), \bibinfo{note}{pMID: 30715845},
  \urlprefix\url{https://doi.org/10.1021/acsnano.8b09059}.

\bibitem[{\citenamefont{Koitzsch et~al.}(2019)\citenamefont{Koitzsch, Pawlik,
  Habenicht, Klaproth, Schuster, B\"uchner, and Knupfer}}]{Koitzsch2019}
\bibinfo{author}{\bibfnamefont{A.}~\bibnamefont{Koitzsch}},
  \bibinfo{author}{\bibfnamefont{A.-S.} \bibnamefont{Pawlik}},
  \bibinfo{author}{\bibfnamefont{C.}~\bibnamefont{Habenicht}},
  \bibinfo{author}{\bibfnamefont{T.}~\bibnamefont{Klaproth}},
  \bibinfo{author}{\bibfnamefont{R.}~\bibnamefont{Schuster}},
  \bibinfo{author}{\bibfnamefont{B.}~\bibnamefont{B\"uchner}},
  \bibnamefont{and} \bibinfo{author}{\bibfnamefont{M.}~\bibnamefont{Knupfer}},
  \bibinfo{journal}{npj 2D Mater. Appl.} \textbf{\bibinfo{volume}{3}},
  \bibinfo{pages}{41} (\bibinfo{year}{2019}),
  \urlprefix\url{https://doi.org/10.1038/s41699-019-0122-6}.

\bibitem[{\citenamefont{Hong et~al.}(2020)\citenamefont{Hong, Senga, Pichler,
  and Suenaga}}]{KSuenaga2020}
\bibinfo{author}{\bibfnamefont{J.}~\bibnamefont{Hong}},
  \bibinfo{author}{\bibfnamefont{R.}~\bibnamefont{Senga}},
  \bibinfo{author}{\bibfnamefont{T.}~\bibnamefont{Pichler}}, \bibnamefont{and}
  \bibinfo{author}{\bibfnamefont{K.}~\bibnamefont{Suenaga}},
  \bibinfo{journal}{Phys. Rev. Lett.} \textbf{\bibinfo{volume}{124}},
  \bibinfo{pages}{087401} (\bibinfo{year}{2020}),
  \urlprefix\url{https://link.aps.org/doi/10.1103/PhysRevLett.124.087401}.

\bibitem[{\citenamefont{Wu et~al.}(2015)\citenamefont{Wu, Qu, and
  MacDonald}}]{AHMacDonald2015a}
\bibinfo{author}{\bibfnamefont{F.}~\bibnamefont{Wu}},
  \bibinfo{author}{\bibfnamefont{F.}~\bibnamefont{Qu}}, \bibnamefont{and}
  \bibinfo{author}{\bibfnamefont{A.~H.} \bibnamefont{MacDonald}},
  \bibinfo{journal}{Phys. Rev. B} \textbf{\bibinfo{volume}{91}},
  \bibinfo{pages}{075310} (\bibinfo{year}{2015}),
  \urlprefix\url{https://link.aps.org/doi/10.1103/PhysRevB.91.075310}.

\bibitem[{\citenamefont{Deilmann and Thygesen}(2019)}]{Deilmann2019}
\bibinfo{author}{\bibfnamefont{T.}~\bibnamefont{Deilmann}} \bibnamefont{and}
  \bibinfo{author}{\bibfnamefont{K.~S.} \bibnamefont{Thygesen}},
  \bibinfo{journal}{2D Materials} \textbf{\bibinfo{volume}{6}},
  \bibinfo{pages}{035003} (\bibinfo{year}{2019}).

\bibitem[{\citenamefont{Bieniek et~al.}(2020)\citenamefont{Bieniek,
  Szulakowska, and Hawrylak}}]{PHawrylak2020}
\bibinfo{author}{\bibfnamefont{M.}~\bibnamefont{Bieniek}},
  \bibinfo{author}{\bibfnamefont{L.}~\bibnamefont{Szulakowska}},
  \bibnamefont{and} \bibinfo{author}{\bibfnamefont{P.}~\bibnamefont{Hawrylak}},
  \bibinfo{journal}{Phys. Rev. B} \textbf{\bibinfo{volume}{101}},
  \bibinfo{pages}{125423} (\bibinfo{year}{2020}).

\bibitem[{\citenamefont{Peng et~al.}(2019)\citenamefont{Peng, Lo, Li, Huang,
  Chen, Lee, Yang, and Cheng}}]{SJCheng2019}
\bibinfo{author}{\bibfnamefont{G.-H.} \bibnamefont{Peng}},
  \bibinfo{author}{\bibfnamefont{P.-Y.} \bibnamefont{Lo}},
  \bibinfo{author}{\bibfnamefont{W.-H.} \bibnamefont{Li}},
  \bibinfo{author}{\bibfnamefont{Y.-C.} \bibnamefont{Huang}},
  \bibinfo{author}{\bibfnamefont{Y.-H.} \bibnamefont{Chen}},
  \bibinfo{author}{\bibfnamefont{C.-H.} \bibnamefont{Lee}},
  \bibinfo{author}{\bibfnamefont{C.-K.} \bibnamefont{Yang}}, \bibnamefont{and}
  \bibinfo{author}{\bibfnamefont{S.-J.} \bibnamefont{Cheng}},
  \bibinfo{journal}{Nano Letters} \textbf{\bibinfo{volume}{19}},
  \bibinfo{pages}{2299} (\bibinfo{year}{2019}), \bibinfo{note}{pMID: 30860847},
  \urlprefix\url{https://doi.org/10.1021/acs.nanolett.8b04786}.

\bibitem[{\citenamefont{Sham and Rice}(1966)}]{LJSham1966}
\bibinfo{author}{\bibfnamefont{L.~J.} \bibnamefont{Sham}} \bibnamefont{and}
  \bibinfo{author}{\bibfnamefont{T.~M.} \bibnamefont{Rice}},
  \bibinfo{journal}{Phys. Rev.} \textbf{\bibinfo{volume}{144}},
  \bibinfo{pages}{708} (\bibinfo{year}{1966}),
  \urlprefix\url{https://link.aps.org/doi/10.1103/PhysRev.144.708}.

\bibitem[{\citenamefont{Hanke and Sham}(1980)}]{LJSham1980}
\bibinfo{author}{\bibfnamefont{W.}~\bibnamefont{Hanke}} \bibnamefont{and}
  \bibinfo{author}{\bibfnamefont{L.~J.} \bibnamefont{Sham}},
  \bibinfo{journal}{Phys. Rev. B} \textbf{\bibinfo{volume}{21}},
  \bibinfo{pages}{4656} (\bibinfo{year}{1980}),
  \urlprefix\url{https://link.aps.org/doi/10.1103/PhysRevB.21.4656}.

\bibitem[{\citenamefont{Rohlfing and Louie}(1998)}]{SGLouie1998}
\bibinfo{author}{\bibfnamefont{M.}~\bibnamefont{Rohlfing}} \bibnamefont{and}
  \bibinfo{author}{\bibfnamefont{S.~G.} \bibnamefont{Louie}},
  \bibinfo{journal}{Phys. Rev. Lett.} \textbf{\bibinfo{volume}{81}},
  \bibinfo{pages}{2312} (\bibinfo{year}{1998}),
  \urlprefix\url{https://link.aps.org/doi/10.1103/PhysRevLett.81.2312}.

\bibitem[{\citenamefont{Vasconcelos et~al.}(2018)\citenamefont{Vasconcelos,
  Bragan{\c{c}}a, Qu, and Fu}}]{vasconcelos2018dark}
\bibinfo{author}{\bibfnamefont{R.}~\bibnamefont{Vasconcelos}},
  \bibinfo{author}{\bibfnamefont{H.}~\bibnamefont{Bragan{\c{c}}a}},
  \bibinfo{author}{\bibfnamefont{F.}~\bibnamefont{Qu}}, \bibnamefont{and}
  \bibinfo{author}{\bibfnamefont{J.}~\bibnamefont{Fu}},
  \bibinfo{journal}{Physical Review B} \textbf{\bibinfo{volume}{98}},
  \bibinfo{pages}{195302} (\bibinfo{year}{2018}).

\bibitem[{Sup()}]{Supplement}
\bibinfo{howpublished}{See Supplemental Material at http://link.aps.org/
  supplemental/xxx/xxx, which includes the detailed information about the first
  principles computation, symmetry analysis, the matrix elements of {\it e-h}
  direct and exchange Coulomb interactions, pseudospin exciton model and the
  theory for the optical polarizations of phonon-assisted indirect
  photoluminescences.}

\bibitem[{\citenamefont{Keldysh}(1979)}]{LVKeldysh1979}
\bibinfo{author}{\bibfnamefont{L.~V.} \bibnamefont{Keldysh}},
  \bibinfo{journal}{JETP Lett.} \textbf{\bibinfo{volume}{29}},
  \bibinfo{pages}{658} (\bibinfo{year}{1979}).

\bibitem[{\citenamefont{Cudazzo et~al.}(2011)\citenamefont{Cudazzo, Tokatly,
  and Rubio}}]{PCudazzo2011}
\bibinfo{author}{\bibfnamefont{P.}~\bibnamefont{Cudazzo}},
  \bibinfo{author}{\bibfnamefont{I.~V.} \bibnamefont{Tokatly}},
  \bibnamefont{and} \bibinfo{author}{\bibfnamefont{A.}~\bibnamefont{Rubio}},
  \bibinfo{journal}{Phys. Rev. B} \textbf{\bibinfo{volume}{84}},
  \bibinfo{pages}{085406} (\bibinfo{year}{2011}),
  \urlprefix\url{https://link.aps.org/doi/10.1103/PhysRevB.84.085406}.

\bibitem[{\citenamefont{Berkelbach et~al.}(2013)\citenamefont{Berkelbach,
  Hybertsen, and Reichman}}]{TCBerkelbach2013}
\bibinfo{author}{\bibfnamefont{T.~C.} \bibnamefont{Berkelbach}},
  \bibinfo{author}{\bibfnamefont{M.~S.} \bibnamefont{Hybertsen}},
  \bibnamefont{and} \bibinfo{author}{\bibfnamefont{D.~R.}
  \bibnamefont{Reichman}}, \bibinfo{journal}{Phys. Rev. B}
  \textbf{\bibinfo{volume}{88}}, \bibinfo{pages}{045318}
  (\bibinfo{year}{2013}),
  \urlprefix\url{https://link.aps.org/doi/10.1103/PhysRevB.88.045318}.

\bibitem[{\citenamefont{Stier et~al.}(2018)\citenamefont{Stier, Wilson,
  Velizhanin, Kono, Xu, and Crooker}}]{AVStier2018}
\bibinfo{author}{\bibfnamefont{A.~V.} \bibnamefont{Stier}},
  \bibinfo{author}{\bibfnamefont{N.~P.} \bibnamefont{Wilson}},
  \bibinfo{author}{\bibfnamefont{K.~A.} \bibnamefont{Velizhanin}},
  \bibinfo{author}{\bibfnamefont{J.}~\bibnamefont{Kono}},
  \bibinfo{author}{\bibfnamefont{X.}~\bibnamefont{Xu}}, \bibnamefont{and}
  \bibinfo{author}{\bibfnamefont{S.~A.} \bibnamefont{Crooker}},
  \bibinfo{journal}{Phys. Rev. Lett.} \textbf{\bibinfo{volume}{120}},
  \bibinfo{pages}{057405} (\bibinfo{year}{2018}),
  \urlprefix\url{https://link.aps.org/doi/10.1103/PhysRevLett.120.057405}.

\bibitem[{\citenamefont{Ridolfi et~al.}(2018)\citenamefont{Ridolfi, Lewenkopf,
  and Pereira}}]{ERidolfi2018}
\bibinfo{author}{\bibfnamefont{E.}~\bibnamefont{Ridolfi}},
  \bibinfo{author}{\bibfnamefont{C.~H.} \bibnamefont{Lewenkopf}},
  \bibnamefont{and} \bibinfo{author}{\bibfnamefont{V.~M.}
  \bibnamefont{Pereira}}, \bibinfo{journal}{Phys. Rev. B}
  \textbf{\bibinfo{volume}{97}}, \bibinfo{pages}{205409}
  (\bibinfo{year}{2018}),
  \urlprefix\url{https://link.aps.org/doi/10.1103/PhysRevB.97.205409}.

\bibitem[{\citenamefont{Trolle et~al.}(2017)\citenamefont{Trolle, Pedersen, and
  V{\'e}niard}}]{MLTrolle2017}
\bibinfo{author}{\bibfnamefont{M.~L.} \bibnamefont{Trolle}},
  \bibinfo{author}{\bibfnamefont{T.~G.} \bibnamefont{Pedersen}},
  \bibnamefont{and}
  \bibinfo{author}{\bibfnamefont{V.}~\bibnamefont{V{\'e}niard}},
  \bibinfo{journal}{Sci. Rep.} \textbf{\bibinfo{volume}{7}},
  \bibinfo{pages}{39844} (\bibinfo{year}{2017}).

\bibitem[{\citenamefont{Ko{\'s}mider et~al.}(2013)\citenamefont{Ko{\'s}mider,
  Gonz{\'a}lez, and Fern{\'a}ndez-Rossier}}]{kosmider2013large}
\bibinfo{author}{\bibfnamefont{K.}~\bibnamefont{Ko{\'s}mider}},
  \bibinfo{author}{\bibfnamefont{J.~W.} \bibnamefont{Gonz{\'a}lez}},
  \bibnamefont{and}
  \bibinfo{author}{\bibfnamefont{J.}~\bibnamefont{Fern{\'a}ndez-Rossier}},
  \bibinfo{journal}{Phys. Rev. B} \textbf{\bibinfo{volume}{88}},
  \bibinfo{pages}{245436} (\bibinfo{year}{2013}).

\bibitem[{\citenamefont{Lado and Fern{\'a}ndez-Rossier}(2016)}]{lado2016landau}
\bibinfo{author}{\bibfnamefont{J.~L.} \bibnamefont{Lado}} \bibnamefont{and}
  \bibinfo{author}{\bibfnamefont{J.}~\bibnamefont{Fern{\'a}ndez-Rossier}},
  \bibinfo{journal}{2D Mater.} \textbf{\bibinfo{volume}{3}},
  \bibinfo{pages}{035023} (\bibinfo{year}{2016}).

\bibitem[{\citenamefont{Mostofi et~al.}(2008)\citenamefont{Mostofi, Yates, Lee,
  Souza, Vanderbilt, and Marzari}}]{Wannier90a}
\bibinfo{author}{\bibfnamefont{A.~A.} \bibnamefont{Mostofi}},
  \bibinfo{author}{\bibfnamefont{J.~R.} \bibnamefont{Yates}},
  \bibinfo{author}{\bibfnamefont{Y.-S.} \bibnamefont{Lee}},
  \bibinfo{author}{\bibfnamefont{I.}~\bibnamefont{Souza}},
  \bibinfo{author}{\bibfnamefont{D.}~\bibnamefont{Vanderbilt}},
  \bibnamefont{and} \bibinfo{author}{\bibfnamefont{N.}~\bibnamefont{Marzari}},
  \bibinfo{journal}{Comput. Phys. Commun.} \textbf{\bibinfo{volume}{178}},
  \bibinfo{pages}{685} (\bibinfo{year}{2008}), ISSN \bibinfo{issn}{0010-4655},
  \urlprefix\url{http://www.sciencedirect.com/science/article/pii/S0010465507004936}.

\bibitem[{\citenamefont{Mostofi et~al.}(2014)\citenamefont{Mostofi, Yates,
  Pizzi, Lee, Souza, Vanderbilt, and Marzari}}]{Wannier90b}
\bibinfo{author}{\bibfnamefont{A.~A.} \bibnamefont{Mostofi}},
  \bibinfo{author}{\bibfnamefont{J.~R.} \bibnamefont{Yates}},
  \bibinfo{author}{\bibfnamefont{G.}~\bibnamefont{Pizzi}},
  \bibinfo{author}{\bibfnamefont{Y.-S.} \bibnamefont{Lee}},
  \bibinfo{author}{\bibfnamefont{I.}~\bibnamefont{Souza}},
  \bibinfo{author}{\bibfnamefont{D.}~\bibnamefont{Vanderbilt}},
  \bibnamefont{and} \bibinfo{author}{\bibfnamefont{N.}~\bibnamefont{Marzari}},
  \bibinfo{journal}{Comput. Phys. Commun.} \textbf{\bibinfo{volume}{185}},
  \bibinfo{pages}{2309} (\bibinfo{year}{2014}), ISSN \bibinfo{issn}{0010-4655},
  \urlprefix\url{http://www.sciencedirect.com/science/article/pii/S001046551400157X}.

\bibitem[{\citenamefont{Kresse and Furthm\"uller}(1996)}]{GKresse1996}
\bibinfo{author}{\bibfnamefont{G.}~\bibnamefont{Kresse}} \bibnamefont{and}
  \bibinfo{author}{\bibfnamefont{J.}~\bibnamefont{Furthm\"uller}},
  \bibinfo{journal}{Phys. Rev. B} \textbf{\bibinfo{volume}{54}},
  \bibinfo{pages}{11169} (\bibinfo{year}{1996}),
  \urlprefix\url{https://link.aps.org/doi/10.1103/PhysRevB.54.11169}.

\bibitem[{\citenamefont{Perdew et~al.}(1996)\citenamefont{Perdew, Burke, and
  Ernzerhof}}]{PBE1996}
\bibinfo{author}{\bibfnamefont{J.~P.} \bibnamefont{Perdew}},
  \bibinfo{author}{\bibfnamefont{K.}~\bibnamefont{Burke}}, \bibnamefont{and}
  \bibinfo{author}{\bibfnamefont{M.}~\bibnamefont{Ernzerhof}},
  \bibinfo{journal}{Phys. Rev. Lett.} \textbf{\bibinfo{volume}{77}},
  \bibinfo{pages}{3865} (\bibinfo{year}{1996}),
  \urlprefix\url{https://link.aps.org/doi/10.1103/PhysRevLett.77.3865}.

\bibitem[{\citenamefont{Qiu et~al.}(2013)\citenamefont{Qiu, da~Jornada, and
  Louie}}]{SGLouie2013a}
\bibinfo{author}{\bibfnamefont{D.~Y.} \bibnamefont{Qiu}},
  \bibinfo{author}{\bibfnamefont{F.~H.} \bibnamefont{da~Jornada}},
  \bibnamefont{and} \bibinfo{author}{\bibfnamefont{S.~G.} \bibnamefont{Louie}},
  \bibinfo{journal}{Phys. Rev. Lett.} \textbf{\bibinfo{volume}{111}},
  \bibinfo{pages}{216805} (\bibinfo{year}{2013}),
  \urlprefix\url{https://link.aps.org/doi/10.1103/PhysRevLett.111.216805}.

\bibitem[{\citenamefont{Chernikov et~al.}(2014)\citenamefont{Chernikov,
  Berkelbach, Hill, Rigosi, Li, Aslan, Reichman, Hybertsen, and
  Heinz}}]{TFHeinz2014a}
\bibinfo{author}{\bibfnamefont{A.}~\bibnamefont{Chernikov}},
  \bibinfo{author}{\bibfnamefont{T.~C.} \bibnamefont{Berkelbach}},
  \bibinfo{author}{\bibfnamefont{H.~M.} \bibnamefont{Hill}},
  \bibinfo{author}{\bibfnamefont{A.}~\bibnamefont{Rigosi}},
  \bibinfo{author}{\bibfnamefont{Y.}~\bibnamefont{Li}},
  \bibinfo{author}{\bibfnamefont{O.~B.} \bibnamefont{Aslan}},
  \bibinfo{author}{\bibfnamefont{D.~R.} \bibnamefont{Reichman}},
  \bibinfo{author}{\bibfnamefont{M.~S.} \bibnamefont{Hybertsen}},
  \bibnamefont{and} \bibinfo{author}{\bibfnamefont{T.~F.} \bibnamefont{Heinz}},
  \bibinfo{journal}{Phys. Rev. Lett.} \textbf{\bibinfo{volume}{113}},
  \bibinfo{pages}{076802} (\bibinfo{year}{2014}),
  \urlprefix\url{https://link.aps.org/doi/10.1103/PhysRevLett.113.076802}.

\bibitem[{\citenamefont{Robert et~al.}(2017)\citenamefont{Robert, Amand, Cadiz,
  Lagarde, Courtade, Manca, Taniguchi, Watanabe, Urbaszek, and
  Marie}}]{CRobert2017}
\bibinfo{author}{\bibfnamefont{C.}~\bibnamefont{Robert}},
  \bibinfo{author}{\bibfnamefont{T.}~\bibnamefont{Amand}},
  \bibinfo{author}{\bibfnamefont{F.}~\bibnamefont{Cadiz}},
  \bibinfo{author}{\bibfnamefont{D.}~\bibnamefont{Lagarde}},
  \bibinfo{author}{\bibfnamefont{E.}~\bibnamefont{Courtade}},
  \bibinfo{author}{\bibfnamefont{M.}~\bibnamefont{Manca}},
  \bibinfo{author}{\bibfnamefont{T.}~\bibnamefont{Taniguchi}},
  \bibinfo{author}{\bibfnamefont{K.}~\bibnamefont{Watanabe}},
  \bibinfo{author}{\bibfnamefont{B.}~\bibnamefont{Urbaszek}}, \bibnamefont{and}
  \bibinfo{author}{\bibfnamefont{X.}~\bibnamefont{Marie}},
  \bibinfo{journal}{Phys. Rev. B} \textbf{\bibinfo{volume}{96}},
  \bibinfo{pages}{155423} (\bibinfo{year}{2017}).

\bibitem[{\citenamefont{Jin et~al.}(2014)\citenamefont{Jin, Li, Mullen, and
  Kim}}]{jin2014intrinsic}
\bibinfo{author}{\bibfnamefont{Z.}~\bibnamefont{Jin}},
  \bibinfo{author}{\bibfnamefont{X.}~\bibnamefont{Li}},
  \bibinfo{author}{\bibfnamefont{J.~T.} \bibnamefont{Mullen}},
  \bibnamefont{and} \bibinfo{author}{\bibfnamefont{K.~W.} \bibnamefont{Kim}},
  \bibinfo{journal}{Physical Review B} \textbf{\bibinfo{volume}{90}},
  \bibinfo{pages}{045422} (\bibinfo{year}{2014}).

\bibitem[{\citenamefont{Chen et~al.}(2020)\citenamefont{Chen, Pieczarka,
  Wurdack, Estrecho, Taniguchi, Watanabe, Yan, Ostrovskaya, and
  Fuhrern}}]{chen2020}
\bibinfo{author}{\bibfnamefont{S.-Y.} \bibnamefont{Chen}},
  \bibinfo{author}{\bibfnamefont{M.}~\bibnamefont{Pieczarka}},
  \bibinfo{author}{\bibfnamefont{M.}~\bibnamefont{Wurdack}},
  \bibinfo{author}{\bibfnamefont{E.}~\bibnamefont{Estrecho}},
  \bibinfo{author}{\bibfnamefont{T.}~\bibnamefont{Taniguchi}},
  \bibinfo{author}{\bibfnamefont{K.}~\bibnamefont{Watanabe}},
  \bibinfo{author}{\bibfnamefont{J.}~\bibnamefont{Yan}},
  \bibinfo{author}{\bibfnamefont{E.~A.} \bibnamefont{Ostrovskaya}},
  \bibnamefont{and} \bibinfo{author}{\bibfnamefont{M.~S.}
  \bibnamefont{Fuhrern}}, \bibinfo{journal}{arXiv:2009.09602
  [cond-mat.mes-hall]}  (\bibinfo{year}{2020}).

\end{thebibliography}

\end{document}